\documentclass[aps, prd, twocolumn, preprintnumbers, nofootinbib, superscriptaddress, balancelastpage, 10pt]{revtex4-2}
\pdfoutput=1 

\usepackage[T1]{fontenc} 
\usepackage{verbatim} 
\usepackage{graphicx}
\usepackage{dcolumn}
\usepackage{bm}
\usepackage{xcolor}
\usepackage[normalem]{ulem}
\usepackage{hyperref}
\usepackage{braket}
\usepackage{amsmath}
\usepackage{amssymb}
\usepackage{bbold}
\usepackage{cleveref}
\usepackage{cancel}
\usepackage{scalerel}
\newcommand{\change}[1]{{\color{black} #1}}

\newcommand{\nn}{\nonumber \\}  

\begin{document}

\title{Muon-Decay Parameters from COHERENT}

\author{Víctor Bres\'o-Pla}
\email{breso@thphys.uni-heidelberg.de}
\address{Institute for Theoretical Physics, Universit\"at Heidelberg, 69120 Heidelberg, Germany}

\author{Sergio Cruz-Alzaga}
\email{alzaga@ific.uv.es}
\address{Instituto de F\'isica Corpuscular (IFIC), CSIC‐Universitat de Val\`encia, E-46980 Paterna, Spain}

\author{Mart\'in Gonz\'alez-Alonso}
\email{martin.gonzalez@ific.uv.es}
\address{Instituto de F\'isica Corpuscular (IFIC), CSIC‐Universitat de Val\`encia, E-46980 Paterna, Spain}

\author{Suraj Prakash}
\email{suraj.prakash@ific.uv.es}
\address{Instituto de F\'isica Corpuscular (IFIC), CSIC‐Universitat de Val\`encia, E-46980 Paterna, Spain}

\begin{abstract}
We demonstrate that measurements of Coherent Elastic Neutrino-Nucleus Scattering (CE$\nu$NS) at spallation sources are valuable probes of muon-decay physics. Using COHERENT data we derive the first direct constraint on the Michel parameters governing 
the $\bar\nu_\mu$ energy distribution. We also discuss future sensitivities, the implications for the Lorentz structure of the interactions mediating muon decay and the application to other neutrino-production mechanisms like pion decay.
\end{abstract}

\maketitle

\section{Introduction}

Four decades after the theoretical prediction of Coherent Elastic Neutrino-Nucleus Scattering (CE$\nu$NS)~\cite{Freedman:1973yd}, this process was first measured by the COHERENT collaboration~\cite{COHERENT:2017ipa} 
using neutrinos produced by the Spallation Neutron Source at the Oak Ridge National Laboratory. At this facility, high-energy protons hit a mercury target to produce $\pi^+$ and $\pi^-$. The latter are absorbed, whereas the former decay at rest into \emph{prompt} neutrinos and antimuons ($\pi^+ \rightarrow \mu^+ \nu_\mu$) which, in turn, produce \emph{delayed} neutrinos and antineutrinos ($\mu^+ \rightarrow e^+ \nu_e \bar{\nu}_\mu$). 
The energy of these (anti)neutrinos is low enough for them to interact coherently with all the nucleons in a target nucleus, leading to an enhancement of the partonic cross section. The COHERENT collaboration has measured this process using different detectors and target materials~\cite{COHERENT:2017ipa,COHERENT:2020iec,COHERENT:2021xmm,COHERENT:2025vuz}, turning a discovery measurement into a precision tool within the Intensity Frontier landscape. Recently, CE$\nu$NS detection has also been reported for reactor and solar neutrinos~\cite{Colaresi:2022obx,PandaX:2024muv,XENON:2024ijk,Ackermann:2025obx}, further expanding the reach of this phenomenon beyond accelerator-based sources. These developments are also relevant for dark matter direct detection, as CE$\nu$NS represents an important background in next-generation experiments~\cite{PandaX:2024muv,XENON:2024ijk,CCM:2021leg,Zema:2020qen}.
 
Overall, COHERENT measurements have been groundbreaking, opening new research directions for applied and fundamental physics. Numerous theoretical studies have analyzed their implications for various aspects of detection physics~\cite{Barranco:2005yy,Scholberg:2005qs,Cadeddu:2017etk,Papoulias:2017qdn,Shoemaker:2017lzs,Liao:2017uzy,Cadeddu:2018dux,AristizabalSierra:2018eqm,Denton:2018xmq,Altmannshofer:2018xyo,Giunti:2019xpr,Coloma:2019mbs,Skiba:2020msb,Hoferichter:2020osn,Miranda:2020tif,AtzoriCorona:2022qrf,DeRomeri:2022twg,Breso-Pla:2023tnz}.
These include studies of the electroweak interaction (extraction of the weak mixing angle), nuclear physics (determination of the neutron skin), electromagnetic neutrino properties, light mediators, and non-standard neutral-current interactions. 
In this letter, we demonstrate for the first time that COHERENT data, and more generally CE$\nu$NS measurements at spallation sources, are also a powerful probe of new physics associated with neutrino production, particularly in the context of muon decay. 

\section{Muon decay parameters}

Non-standard effects can be studied in a model-independent way by focusing on the appropriate Effective Field Theory. The production of (anti)neutrinos via muon decay occurs at energies well below the GeV scale. Hence, the leading terms of the most general effective Lagrangian describing the process $\mu^+\to e^+\nu_e\bar\nu_\mu$ compatible with Lorentz symmetry are
\begin{eqnarray}\label{eq:Lag-muon-decay}
    \hspace{-0.3cm}
    \mathcal{L}
    &=& 
    -
    \frac{4\,G_F}{\sqrt{2}}\sum_{X,\eta,\epsilon}\,
    g^{X}_{\epsilon \eta}(\bar{e}_{\epsilon}\Gamma^X (\nu_{e})_{\rho})((\bar{\nu}_{\mu})_{\gamma} \Gamma_X \mu_\eta)\,+\,h.c.,
\end{eqnarray}
where $G_F\approx 1.166\times 10^{-5}$~GeV$^{-2}$~\cite{ParticleDataGroup:2024cfk} is the Fermi constant, 
$\eta,\epsilon,\rho,\gamma$ are the field chiralities\footnote{For a given Lorentz structure, only two of the chiralities are independent, which is why the sum runs only over $\eta$ and $\epsilon$. 
} and $X$ describes the different Lorentz structures: $\Gamma_X=\mathbb{1},\,\gamma_\eta,\,\sigma_{\eta\rho}/\sqrt{2}$, for $X=S,V,T$, respectively. 

The tensor couplings $g_{LL}^{T}$ and $g_{RR}^{T}$ vanish due to Fierz identities, reducing the muon-decay effective Lagrangian to 10 distinct terms. Moreover, because the Fermi constant is extracted from the muon decay rate, the (complex) parameters $g^{X}_{\epsilon \eta}$ are not independent, but must satisfy the following normalization condition~\cite{Fetscher:1986uj,Fetscher:1994nv,MuonDecayParameters:2024cfk}
\begin{eqnarray}\label{eq:muon_decay_norm}
    \sum_{\epsilon,\eta}\left[
    \frac{1}{4}\, |g^{S}_{\epsilon\eta}|^2 
    \,+\, |g^{V}_{\epsilon\eta}|^2
    \,+\,3 \,
    |g^{T}_{\epsilon\eta}|^2 \right] = 1.
\end{eqnarray}
which sets an upper limit on the size of 
the Wilson Coefficients $g^{X}_{\epsilon \eta}$. 
In the SM limit $g^V_{LL} \rightarrow 1$, while all other couplings vanish.

Measurements based on the detection of the emitted electron or positron have long been used to constrain the coupling parameters $g^{X}_{\epsilon \eta}$~\cite{PhysRevD.37.587,Gagliardi:2005fg,TWIST:2008myj,TWIST:2011aa}. While certain parameter combinations are tightly constrained, others depend on polarization measurements, which generally lead to weaker bounds. Notably, these observables are unable to disentangle the combination $|g_{LL}^{V}|^2 + |g_{LL}^{S}|^2/4$, a point to which we will return later. For an up-to-date summary of existing constraints, we refer the reader to the review by the Particle Data Group (PDG)~\cite{MuonDecayParameters:2024cfk}. In this work, we focus on measurements of the (anti)neutrinos emitted in muon decay instead.

The most general form of the differential energy distribution for electron neutrinos in $\mu^+\to e^+\nu_e\bar\nu_\mu$ can be parametrized as
\begin{eqnarray}\label{eq:DifferentialWidthNeutrino}
    \frac{d\Gamma_{\nu_{{}_H}}}{dE_\nu} 
    &=& 
    \frac{24\Gamma_\mu}{m_\mu} P_{\nu_{{}_H}} \bigg[ y^2 \left(1-y\right) + \frac{8}{9} w_{\nu_{{}_H}} y^2 \left(y-\frac{3}{4}\right) \bigg],~~
\end{eqnarray}
where $H=L,R$ for left- or right-handed neutrinos, $y=2E_\nu / m_\mu$ with $E_\nu$ and $m_\mu$ being the neutrino energy and muon mass respectively, and $\Gamma_\mu$ is the muon decay width. $P_{\nu_{{}_H}}$ denotes the probability for the emission of a neutrino with chirality $H (=L,R)$ and $w_{\nu_{{}_H}}$ is the corresponding spectrum shape parameter.\footnote{The probability $P_x$ is denoted by $Q_x$ in the muon-decay literature. Here we use $P_x$ because $Q$ is used to denote the nuclear weak charge, which plays a central role in CE$\nu$NS observables.} These parameters are the analogue of the Michel parameters for the electron energy distribution~\cite{Michel:1949qe}.

Concerning the differential energy distribution for muon  antineutrinos in $\mu^+\to e^+\nu_e\bar\nu_\mu$, we find that the most general expression has a similar yet distinct parametric form, namely
\begin{eqnarray}\label{eq:DifferentialWidthAntineutrino}
    \frac{d\Gamma_{\bar\nu_{{}_H}}}{dE_{\bar \nu}} 
    &=& 
    \frac{24\Gamma_\mu}{m_\mu} P_{\bar{\nu}_{{}_H}} \bigg[y^2\!\left(\frac{1}{2}-\frac{y}{3}\right) + \frac{8}{9} w_{\bar{\nu}_{{}_H}} y^2\!\left(\frac{3}{4} - y \right)\!\bigg],~~~~
\end{eqnarray}
where $y=2E_{\bar\nu} / m_\mu$, $H=L,R$ refers to the chirality and $P_{\bar\nu_{{}_H}}$ and $w_{\bar\nu_{{}_H}}$ are the corresponding probability and shape parameters. \change{We note that small corrections due to electron and neutrino masses are neglected in Eqs.~\eqref{eq:DifferentialWidthNeutrino} and~\eqref{eq:DifferentialWidthAntineutrino}.}

For left-handed (anti)neutrinos, the Michel parameters $P_x$, $w_x$ are defined in terms of the $g^X_{\epsilon\eta}$ coefficients as
{\small
\begin{align}\label{eq:P-w-L-PDG}
    P_{\nu_L} 
    &=
    |g^{V}_{LL}|^2 + |g^{V}_{LR}|^2 + \frac{1}{4} |g^{S}_{RR}|^2 + \frac{1}{4}|g^S_{RL}|^2 + 3|g^{T}_{RL}|^2~,  \nn
    P_{\nu_L} w_{\nu_L}
    &= \frac{3}{16} 
    \left(4 |g^{V}_{LR}|^2 + |g^{S}_{RR}|^2 + \left| g^{S}_{RL} + 2\,g^{T}_{RL}\right|^2 \right)~,\nn
    P_{\bar{\nu}_L}
    &=
    |g^{V}_{LL}|^2 + |g^{V}_{RL}|^2 + \frac{1}{4} |g^{S}_{RR}|^2 + \frac{1}{4} |g^{S}_{LR}|^2 + 3 |g^{T}_{LR}|^2~, \nn
    P_{\bar{\nu}_L} w_{\bar{\nu}_L}
    &= \frac{3}{16}
     \left(|g^{S}_{RR}|^2 + |g^{S}_{LR}|^2 - 4 |g^{T}_{LR}|^2\right)~.
\end{align}}
The corresponding expressions for right-handed (anti)neutrinos are obtained by interchanging $L \leftrightarrow R$ in each of the above expressions. Let us note that $P_{\nu_L} + P_{\nu_R} = P_{\bar{\nu}_L} + P_{\bar{\nu}_R} = 1 $. \change{In the SM, the expression reduces to $P_{\nu_L} = P_{\bar{\nu}_L} = 1$ due to the absence of right-handed neutrino fields.} 

Our expressions for left-handed neutrinos agree with Ref.~\cite{Fetscher:1994nv}\footnote{We note that a $-8/9$ factor is missing in 
Eq.~(57.8) of the PDG review on muon-decay parameters~\cite{MuonDecayParameters:2024cfk} (as well as in previous editions), i.e., in the neutrino energy distribution, {\it cf}. Eq.~\eqref{eq:DifferentialWidthNeutrino}.} while, to the best of our knowledge, the results for antineutrinos are presented here for the first time.

\begin{figure}[htb]
    \centering
    \includegraphics[width=0.95\linewidth]{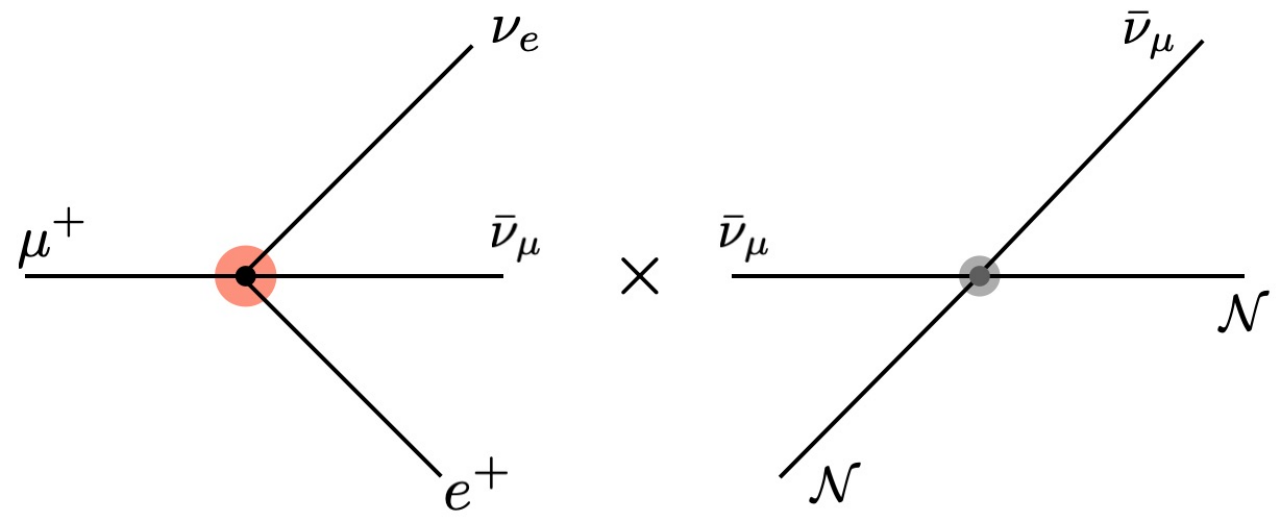}\hfill
    \caption{   
    Feynman diagram for $\mu^+$ decay and the subsequent CE$\nu$NS detection of the emitted muon antineutrino, providing access to its energy distribution.
    }
    \label{fig:diagrams}
\end{figure}

Remarkably, only one experimental extraction of the neutrino Michel parameters has been performed, conducted by KARMEN in 1998~\cite{PhysRevLett.81.520} (following the proposal in Ref.~\cite{Fetscher:1994nv}). Even more strikingly, no experimental extractions of the antineutrino Michel parameters exist. In the remainder of this letter, we present the first such extraction using COHERENT data, 
as illustrated in Fig.~\ref{fig:diagrams}.

\section{COHERENT event rate}

The number of prompt and delayed neutrinos detected by the COHERENT collaboration, produced in $\pi^+\to\mu^+\nu_\mu$ and $\mu^+\to e^+\nu_e\bar\nu_\mu$ decays respectively, can be computed as a convolution of the associated differential (anti)neutrino fluxes and the CE$\nu$NS cross section, i.e., 
\begin{align}\label{eq:FluxTimesXsection}
    \frac{dN^{\rm pr./del.}}{dT} 
    &= \sum_f
    N_T \int dE_\nu \frac{d\phi_f}{dE_\nu}
    \frac{d\sigma_f}{dT}~,
\end{align}
where the sum goes over $f=\nu_e,\bar\nu_\mu$ for delayed (anti)neutrinos and $f=\nu_\mu$ for the prompt component, $T$ is the nuclear recoil energy, $N_T$ is the number of target particles, and $E_\nu$ refers to the (anti)neutrino energy.

The differential fluxes for muon decay are given by the $\nu_L$ and $\bar\nu_L$ distributions in Eqs.~\eqref{eq:DifferentialWidthNeutrino}-\eqref{eq:DifferentialWidthAntineutrino} times $N_f / (4\pi L^2)$, where $L$ is the distance to the detector and $N_f$ is the total number of emitted (anti)neutrinos, $f=\nu_e,\bar{\nu}_{\mu}$. 
On the other hand, for prompt events (pion decay) and CE$\nu$NS detection we will assume SM interactions, since our goal is to study, for the first time, the sensitivity of COHERENT measurements to nonstandard effects in muon decay.\footnote{See Sec.~\ref{sec:pion} for a discussion of nonstandard effects in pion decay.}

Convolving these fluxes with the CE$\nu$NS cross section and summing over $f = \nu_\mu ~(f=\bar{\nu}_{\mu}, \nu_e)$ we obtain the final event rate for prompt (delayed) events.  
We find that the final result can be conveniently re-written as follows
\begin{align}\label{eq:result}
    \frac{dN^{\rm pr./del.}}{dT} &= \sum_f N_T \int dE_\nu \,x_f\,\frac{d\phi^{\rm SM}_f}{dE_\nu} \frac{d\sigma^{\rm SM}}{dT}~,
\end{align}
where $d\phi^{\rm SM}_f/dE_\nu$ \change{are the usual SM fluxes and $d\sigma^{\rm SM}/dT$ is the SM CE$\nu$NS cross section 
\cite{Freedman:1973yd}
\begin{align}
\frac{d\sigma^{\rm SM}}{dT} \approx \frac{m_{\cal N}G_F^2({\cal F}(T))^2}{4\,\pi}    \bigg ( 1   -  { m_{\cal N} \,T \over 2  E_\nu^2} \bigg )  Q_{\rm SM}^2~,
\end{align}
where $m_{\cal N}$ is the nuclear mass and ${\cal F}(T)$ is the nuclear form factor.} 
Nonstandard effects are captured by the $x_f$ parameters, which are given by
\begin{align}\label{eq:x2-in-terms-of-Q-w}
    x_{\bar\nu_\mu} &= P_{\bar\nu_L} - \frac{4}{3}\,P_{\bar\nu_L} w_{\bar\nu_L} + \frac{4}{3}\,P_{\nu_L} w_{\nu_L} ~,\nn
    x_{\nu_e} &= P_{\nu_L} - \frac{4}{3}\,P_{\nu_L} w_{\nu_L}  + \frac{4}{3}\,P_{\bar\nu_L} w_{\bar\nu_L}~,
\end{align}
and $x_{\nu_\mu}=1$ for prompt events. Let us note that the individual delayed fluxes are not proportional to their SM values, i.e., $d\phi_f/dE_\nu\neq x_f\,d\phi^{\rm SM}_f / dE_\nu$ for each $f=\bar{\nu}_{\mu}, \nu_e$ separately, as clearly seen in Eqs.~\eqref{eq:DifferentialWidthNeutrino}-\eqref{eq:DifferentialWidthAntineutrino}. Only after convolving them with the \change{SM cross section} and summing both of them can the result be written as a rescaled sum of the SM fluxes times the cross section, as in Eq.~\eqref{eq:result}. This is why $x_{\bar\nu_\mu}$ contains some of the events mediated by $\nu_e$ (and likewise for $x_{\nu_e}$ and $\bar\nu_\mu$).

Thus, the study of New Physics effects in muon decay can be implemented in existing analyses (which assume SM production) in a trivial way, since the $x_f$ factors can be absorbed in a flavor-dependent {\it effective} weak nuclear charge, namely:
 \begin{align}\label{eq:EffectiveQ}
  \widetilde{Q}_f^2 &= Q_{\rm SM}^2~x_f~,~~(f = \nu_\mu,\bar\nu_\mu, \nu_e)~,
\end{align}
or equivalently, in an {\it effective} cross section, $d\tilde\sigma_f/dT=x_f\,d\sigma^{\rm SM}/dT$. We stress that the effective charges (or cross sections) for $\nu_\mu$ and $\bar\nu_\mu$ are {\it not} equal. 
In principle one can disentangle the effective weak nuclear charges  $\widetilde{Q}_{\bar\nu_\mu}$ and $\widetilde{Q}_{\nu_e}$ (i.e., $x_{\bar\nu_\mu}$ and $x_{\nu_e}$) because the recoil dependence is different for the delayed neutrino and antineutrino SM contributions. However, we show below that current data do not allow for an accurate separation, although one can precisely extract one specific combination of them. This limit on the charges will translate into a constraint on a specific combination of the Michel parameters $P_{\nu_L}$, $P_{\bar\nu_L}$, $w_{\nu_L}$ and $w_{\bar\nu_L}$ using Eq.~\eqref{eq:x2-in-terms-of-Q-w}.

\section{Numerical Analysis}

We use the two most precise CE$\nu$NS measurements currently available, conducted by the COHERENT collaboration with liquid argon (LAr) and cesium iodide (CsI) targets~\cite{COHERENT:2020iec,COHERENT:2020ybo,COHERENT:2021xmm}. For these measurements, double-binned distributions in nuclear recoil and time are available, with $4\times10$ bins in LAr and $52\times12$ bins in CsI.

For a detailed discussion of our analysis of the 
current data, we refer the reader to Ref.~\cite{Breso-Pla:2023tnz}, where the technical aspects of our statistical treatment are presented. This includes the incorporation of various backgrounds and systematic effects following the COHERENT prescription. \change{The crucial point is that Ref.~\cite{Breso-Pla:2023tnz} carried out the simultaneous extraction of the three flavor-dependent effective charges ($\widetilde{Q}_{\nu_e}^2$, $\widetilde{Q}_{\nu_\mu}^2$, and $\widetilde{Q}_{\bar\nu_\mu}^2$).  
In this work, we use Eq.~\eqref{eq:EffectiveQ} to apply those results to a scenario where New Physics is present only in muon decay, in order to extract the $x_f$ parameters.} 

The future of CE$\nu$NS studies at spallation sources is promising, with several ongoing and proposed efforts aimed at refining current measurements using a variety of targets and facilities such as the SNS and the European Spallation Source~\change{\cite{Baxter:2019mcx,Barbeau:2021exu,COHERENT:2022nrm,Abele:2022iml,COHERENT:2023sol}}. 
A detailed comparison of these experiments is beyond the scope of this work. Instead, we take the CENNS-750 experiment, an extension of COHERENT featuring a 610 kg liquid argon detector at the SNS~\cite{COHERENT:2022nrm,Jeong_NuFact2023}
as a valuable reference point for estimating the potential reduction in statistical uncertainties. 
The projections presented below illustrate the substantial opportunities for improvement, which must ultimately be refined through dedicated studies of systematic uncertainties and background contributions.

Using the latest COHERENT data with CsI and LAr detectors~\cite{COHERENT:2020iec,COHERENT:2021xmm}, we obtain $x_{\bar\nu_\mu}=-1.5(1.3)$ and $x_{\nu_e}=4.6(2.1)$ with a large correlation ($\rho = -0.98$). 
The latter implies there is a strongly constrained combination, namely
\begin{eqnarray}
0.84\,x_{\bar\nu_\mu} + 0.54\,x_{\nu_e} &= 1.25(21)~,
\end{eqnarray}
at one sigma. This bound can be expressed in terms of the Michel parameters as follows
\begin{widetext} 
\begin{eqnarray}\label{eq:mainresult}
    & 0.54\, P_{\nu_L} + 0.84\, P_{\bar\nu_L} + 0.40\, \left(P_{\nu_L}w_{\nu_L} - P_{\bar\nu_L}w_{\bar\nu_L}\right)  = 1.25\pm 0.21,
\end{eqnarray}
\end{widetext}
at one sigma. This result is the first direct experimental constraint for the muon antineutrino Michel parameters ($P_{\bar\nu_L}$ and $w_{\bar\nu_L}$) in Eq.~\eqref{eq:DifferentialWidthAntineutrino}. Moreover, it is only the second constraint available for the electron neutrino parameters ($P_{\nu_L}$ and $w_{\nu_L}$) in Eq.~\eqref{eq:DifferentialWidthNeutrino} after the determination by KARMEN~\cite{Fetscher:1994nv,PhysRevLett.81.520}.

\begin{figure}
    \centering
    \includegraphics[width=0.95\linewidth]{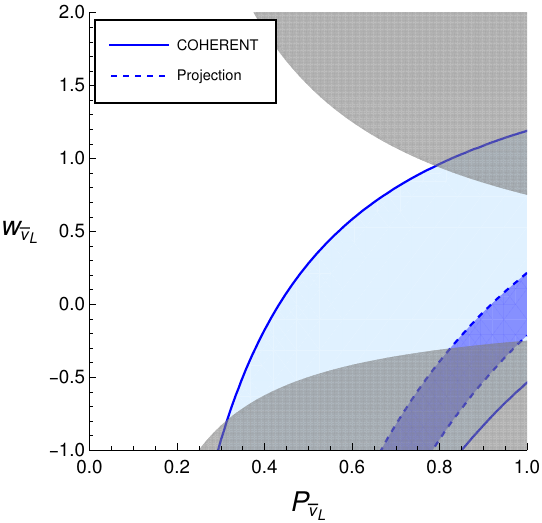}
    \caption{Allowed region (90\% CL) for the $\bar\nu_\mu$ Michel parameters, $P_{\bar\nu_L}$ and $w_{\bar\nu_L}$, {\it cf.} Eq.~\eqref{eq:DifferentialWidthAntineutrino}.
    The use of $P_{\nu_L}=1$ and $w_{\nu_L}=0$ (SM values) or the intervals $P_{\nu_L} \,>\, 0.92$ and $w_{\nu_L} < 0.12$~\cite{PhysRevLett.81.520}, leads to negligible differences in the figure. The solid (dashed) line is the current (projected) bound. The region with unphysical values for $P_{\bar\nu_L}$ and $w_{\bar\nu_L}$ is grayed out.
}
    \label{fig:PvLb-wvLb-1}
\end{figure}

In Fig.~\ref{fig:PvLb-wvLb-1} we show the allowed region for the antineutrino parameters $P_{\bar\nu_L}$ and $w_{\bar\nu_L}$ based on Eq.~\eqref{eq:mainresult} and projected COHERENT data. The choice of $P_{\nu_L}=1$ and $w_{\nu_L}=0$ (SM values) or the currently allowed intervals, $P_{\nu_L} \,>\, 0.92$ and $w_{\nu_L} < 0.12$~\cite{PhysRevLett.81.520}, has a negligible impact on the regions in Fig.~\ref{fig:PvLb-wvLb-1}.

Finally, the one-at-a-time (1-sigma) values are
\begin{align}
P_{\nu_L}& = 0.98\left(^{+02}_{-37}\right)~,\qquad w_{\nu_L} = 0.00\left(^{+23}_{-00}\right)~,\nn
P_{\bar\nu_L}& = 0.90\left(^{+10}_{-24}\right)~,\qquad w_{\bar\nu_L} = 0.51 \left(^{+24}_{-51}\right)~.
\end{align}

\begin{figure}[!htb]
    \centering    \includegraphics[width=0.899\linewidth]{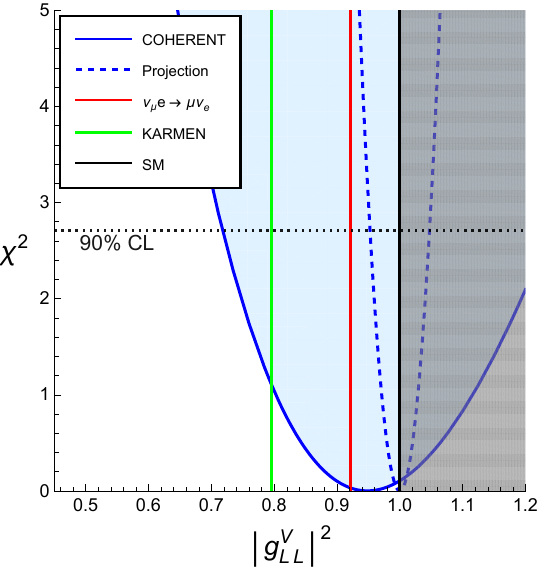}\hfill
    \caption{  
    $\Delta\chi^2$ function for the $g^V_{LL}$ Wilson Coefficient, assuming only $g^V_{LL}$ and $g^S_{LL}$ are present. 
    The solid (dashed) line is the current (projected) bound. 
    The black line refers to the SM limit and the gray region to the excluded region, Eq.~\eqref{eq:muon_decay_norm}. \change{We also include the lower $90\%$ CL bound from inverse muon decay (red) and KARMEN (green). Their upper bound is the SM value, i.e., the unity.}
    }
    \label{fig:hVLL-hSLL}
\end{figure}

The determinations above offer novel information about the structure of the interactions mediating muon decay. Indeed, using Eqs.~\eqref{eq:P-w-L-PDG} we can write the Michel parameters $P_x$ and $w_x$ in terms of the Wilson Coefficients $g^X_{\epsilon\eta}$ of the muon-decay Lagrangian. 
To illustrate this, let us consider the scenario where only $g^V_{LL}$ and $g^S_{LL}$ are present. This case is particularly interesting because traditional muon-decay measurements (where only the electron/positron is detected) cannot provide any constraint.\footnote{They are sensitive to the combination $|g^V_{LL}|^2+|g^S_{LL}|^2/4$, which is fixed to one by the normalization condition, Eq.~\eqref{eq:muon_decay_norm}.} In the past, this flat direction was broken using inverse muon decay data ($\nu_\mu e\to \mu \nu_e$)~\change{~\cite{MISHRA1990170,CHARM-II:1995xfh}}, which is governed by the same interactions. Our results show that COHERENT also breaks this flat direction, since it is sensitive to the vector coupling, $g^V_{LL}$, but not to the scalar one, $g^S_{LL}$, which involves right-handed neutrino fields. Indeed COHERENT data implies
\begin{align}
    |g^V_{LL}|^2 &=0.95^{+0.05}_{-0.15}~,
\end{align}
at one sigma, in agreement with one, which is the SM value. This result, along with our projection, is shown in Fig.~\ref{fig:hVLL-hSLL}. For comparison, inverse muon decay gives $|g^V_{LL}|^2>0.88$\change{~\cite{MISHRA1990170,CHARM-II:1995xfh}}, and the KARMEN determination of $P_{\nu_L}$~\cite{PhysRevLett.81.520} implies $|g^V_{LL}|^2>0.80$, both at 90\% CL. They are more stringent than the current bound by COHERENT, but could be improved by future CE$\nu$NS measurements, as shown in Fig.~\ref{fig:hVLL-hSLL}.

The implications for other Wilson Coefficients, $g^X_{\epsilon\eta}$, and more general scenarios, will be studied in Ref.~\cite{CompanionPaper}. \change{Among the nonstandard couplings, tensor coefficients present the highest sensitivity. This includes $g^V_{RR}$ and $g^S_{LL}$, which can be probed indirectly through the normalization condition, Eq.~\eqref{eq:muon_decay_norm}.} 
\section{Pion decay}
\label{sec:pion}
The same reasoning applies when studying non-standard effects in pion decay ($\pi^+\to\mu^+\nu_\mu$). To illustrate this, let us consider only pseudoscalar couplings:
\begin{eqnarray}\label{eq.Lag-pion-decay}
        \Delta\mathcal{L}
        = \sqrt{2}\, G_F\, V_{ud}\,
        &&\left\{ \,\epsilon_P(\bar{u}\gamma^5 d)(\bar\mu\,  P_L\, \nu_\mu)\right.\nn
        && \left. \!+\,\widetilde{\epsilon}_P(\bar{u}\gamma^5 d)(\bar\mu\,P_R\, \nu_\mu) \right\} + ~h.c.,
\end{eqnarray}
where $V_{ud}$ is the (1,1) element of the Cabibbo--Kobayashi--Maskawa~(CKM) matrix.

In this case, the neutrino energy spectrum is fixed by the two-body kinematics, but the overall rate is not. The effect, encoded in the $x_{\nu_\mu}$ 
factor in Eq.~\eqref{eq:result}, is given by
\begin{eqnarray}
    x_{\nu_\mu}\approx 1-|\widetilde\epsilon_P|^{\,2} \frac{m^4_{\pi^\pm}}{m_{\mu}^2(m_u+m_d)^2}~,
\end{eqnarray}
which represents the probability that the pion emits a left-handed neutrino.

Using current (projected) COHERENT data we obtain $x_{\mu} = 1.41\pm 0.32$ ($x_{\mu} = 1.00\pm 0.05$) at 1$\sigma$, leading to 
\begin{eqnarray}
\widetilde\epsilon_P = 0.000\pm 0.012 ~(0.008)~,
\end{eqnarray}
corresponding to an effective New Physics scale of $2.3~(2.7)$ TeV. This low-energy constraint is unique: pseudoscalar interactions do not contribute to nuclear beta decay at leading order in recoil~\cite{Jackson:1957zz}, and comparing the measured pion decay width to the Standard Model prediction—calculated using the lattice pion decay constant~\cite{FlavourLatticeAveragingGroupFLAG:2024oxs}— only constrains a specific combination of $\epsilon_P$ and $\widetilde\epsilon_P$. \change{The same applies to the $R_e/R_{\mu}$ ratio, which also  involves couplings to electron fields.} 
For a discussion of LHC constraints on the underlying high-energy contact interactions, which lie in the $0.001$ range depending on the validity of the Effective Field Theory expansion, see Ref.~\cite{Falkowski:2017pss}.

\section{Conclusions and outlook}

Let us recap our results, add some additional comments and discuss possible extensions of our work.\

The first CE$\nu$NS measurements have been widely used to study various aspects of neutrino detection. In this letter, we have highlighted that COHERENT, and more generally CE$\nu$NS measurements at spallation sources, are also a powerful probe of new physics associated with neutrino production. In particular, we have focused on muon decay ($\mu^+\to e^+\nu_e\bar\nu_\mu$). Using the most precise COHERENT data we have carried out the first extraction of the $\bar\nu_\mu$ Michel parameters, describing the associated energy distribution. We have also explored the implications for the underlying structure of the interactions mediating muon decay, the effects on pion decay and the impact of future CE$\nu$NS measurements at spallation sources.

We encourage the COHERENT collaboration to analyze their future measurements within this framework ---requiring only the use of flavor-dependent weak nuclear charges--- and to carry out the extraction of muon-decay parameters, which should be incorporated into future editions of the PDG review~\cite{MuonDecayParameters:2024cfk}.

A natural extension of the results in this letter is to consider a scenario where nonstandard effects arise simultaneously in muon decay, pion decay, CE$\nu$NS detection, and a general flavor structure is allowed. In particular, the latter introduces muon decay channels involving  ``wrong-flavor'' neutrinos, such as $\mu^+\to e^+\nu_\tau\bar\nu_\tau$. Addressing this more general case requires a more sophisticated theoretical framework, which we will present in a forthcoming, more technical article~\cite{CompanionPaper}. Let us mention that the parametric form of the differential distributions in (anti)neutrino energy in Eqs.~\eqref{eq:DifferentialWidthAntineutrino} remains valid in the presence of general lepton-flavor violating (LFV) couplings, where the Michel parameters become functions of the new couplings as well. Consequently, our main result in Eq.~\eqref{eq:mainresult} still applies, but with broader implications. Notably, it provides unique bounds on high-energy LFV contact interactions, which are inaccessible to traditional electroweak precision observables, LFV probes, or collider searches, and are relevant for concrete New Physics models and Effective Field Theories valid at collider scales, such as the SMEFT~\cite{Buchmuller:1985jz,Grzadkowski:2010es,Falkowski:2023hsg}.

Finally, our approach can also be applied to the recent CE$\nu$NS measurements with reactor and solar neutrinos~\cite{Colaresi:2022obx,XENON:2024ijk,PandaX:2024muv,Ackermann:2025obx} to probe nonstandard couplings in the corresponding production processes. 
These experiments require a dedicated study due to the complexity of the production mechanisms, which involve a large number of nuclear beta decays. Additionally, for solar neutrinos, oscillation and matter effects must also 
be taken into account. 

Overall, the results obtained in this work, along with the new directions they open, broaden the applications of CE$\nu$NS measurements in fundamental physics.
\vspace{0.8cm}

\acknowledgments
We thank Adam Falkowski and Toni Pich for useful comments and discussions. 
This work has been supported by MCIN/AEI/10.13039/501100011033 (grants PID2020-114473GB-I00 and PID2023-146220NB-I00) and by MICIU/AEI/10.13039/501100011033 and European Union NextGenerationEU/PRTR (grant CNS2022-135595). VB is supported by the DFG under grant 396021762 – TRR 257: \textit{Particle Physics Phenomenology after the Higgs Discovery} and the BMBF Junior \textit{Group Generative Precision Networks for Particle Physics} (DLR 01IS22079).

\section*{Data Availability}

The data that support the findings of
this Letter are openly available \cite{COHERENT:2021xmm,COHERENT:2020ybo}.

\bibliographystyle{apsrev4-1}
\bibliography{muDecay_COHERENT}

\end{document}